\documentclass{egpubl}
\usepackage{eurovis2026}

\SpecialIssuePaper         

\CGFccby

\usepackage[T1]{fontenc}
\usepackage{dfadobe}  

\BibtexOrBiblatex
\usepackage[backend=biber,bibstyle=EG,citestyle=alphabetic,maxcitenames=2,
  mincitenames=1]{biblatex} 
\DeclareCiteCommand{\citet}
  {\usebibmacro{prenote}}
  {%
    \begingroup

    \printnames{labelname}%
    \endgroup
    \addspace
    \mkbibbrackets{\usebibmacro{cite}}%
  }
  {\multicitedelim}
  {\usebibmacro{postnote}}

\electronicVersion
\ifpdf \usepackage[pdftex]{graphicx} \pdfcompresslevel=9
\else \usepackage[dvips]{graphicx} \fi

\usepackage{egweblnk}
\usepackage{float}
\usepackage[skins]{tcolorbox}
\usepackage[HTML,table,xcdraw]{xcolor}

\usepackage{tabularx}
\usepackage{booktabs}
\usepackage{enumitem}
\usepackage{multicol}
\usepackage{longtable}
\usepackage{amsmath}
\usepackage{multirow}
\usepackage{hyperref}
\hypersetup{
        colorlinks=true,
        allcolors=black
    }
\usepackage[normalem]{ulem}

\newcommand{\acro}[1]{#1}

\definecolor{aiAgentColor}{HTML}{0092E9} 
\definecolor{humanAgentColor}{HTML}{C96A20} 
\definecolor{sharedGoalsColor}{HTML}{7C546E} 
\definecolor{miImpactColor}{HTML}{3A8435} 
\definecolor{miPrinciplesColor}{HTML}{C00000} 
\definecolor{evaluationColor}{HTML}{156082} 
\definecolor{levelOfAutoColor}{HTML}{404040} 

\newcommand{\autol}[1]{\leavevmode{\color{levelOfAutoColor}{#1}}}
\newcommand{\sharedtask}[1]{\leavevmode{\color{sharedGoalsColor}{#1}}}
\newcommand{\impact}[1]{\leavevmode{\color{miImpactColor}{#1}}}
\newcommand{\eval}[1]{\leavevmode{\color{evaluationColor}{#1}}}
\newcommand{\principle}[1]{\leavevmode{\color{miPrinciplesColor}{#1}}}

\newif\ifrevisions
\revisionsfalse 

\ifrevisions 
    \long\def\added#1{{\color{blue}#1}} 
    \long\def\deleted#1{{\color{red}\sout{#1}}} 
\else 
    \long\def\added#1{#1} 
    \long\def\deleted#1{} 
\fi 

\title[A Review of Mixed Initiative Visual Analytics]%
      {A Scoping Review of Mixed Initiative Visual Analytics \\ in the Automation Renaissance}

\author[Monadjemi et al.]
{\parbox{\textwidth}{\centering 
        Shayan Monadjemi\thanks{monadjemis@ornl.gov}$^{1}$\orcid{0000-0002-9385-5969},
        Yuhan Guo$^{2}$\orcid{0009-0004-3857-7486}, 
        Kai Xu$^{3}$\orcid{0000-0003-2242-5440}, 
        Alex Endert$^{4}$\orcid{0000-0002-6914-610X}, and 
        Anamaria Crisan\thanks{ana.crisan@uwaterloo.ca}$^{5}$\orcid{0000-0003-3445-3414}
        }
        \\
{\parbox{\textwidth}{\centering 
        $^1$Oak Ridge National Laboratory, USA
        \hspace{2em}
        $^2$Peking University, China
        \hspace{2em}
        $^3$University of Nottingham, UK \\
        $^4$Georgia Institute of Technology, USA
        \hspace{2em}
        $^5$University of Waterloo, Canada
       }
}
}

\makeatletter
\g@addto@macro\ps@titlepage{%
  \def\@oddfoot{%
    {\tiny\raisebox{\z@}[8pt][1pt]{\parbox[t]{20pc}{\sloppy
      \p@copyrightTextTitPag}}}%
    \hfill
    {\tiny\raisebox{\z@}[8pt][1pt]{\parbox[t]{20pc}{%
            Notice: This manuscript has been authored by UT-Battelle, LLC, under contract DE-AC05-00OR22725 with the US Department of Energy (DOE). The US government retains and the publisher, by accepting the article for publication, acknowledges that the US government retains a nonexclusive, paid-up, irrevocable, worldwide license to publish or reproduce the published form of this manuscript, or allow others to do so, for US government purposes. DOE will provide public access to these results of federally sponsored research in accordance with the DOE Public Access Plan ( https://www.energy.gov/doe-public-access-plan ).
      }}}%
  }%
}
\makeatother

\begin{document}

\teaser{
  \includegraphics[clip, trim=0.5cm 3.5cm 0.5cm 3.65cm, width=0.9\textwidth]{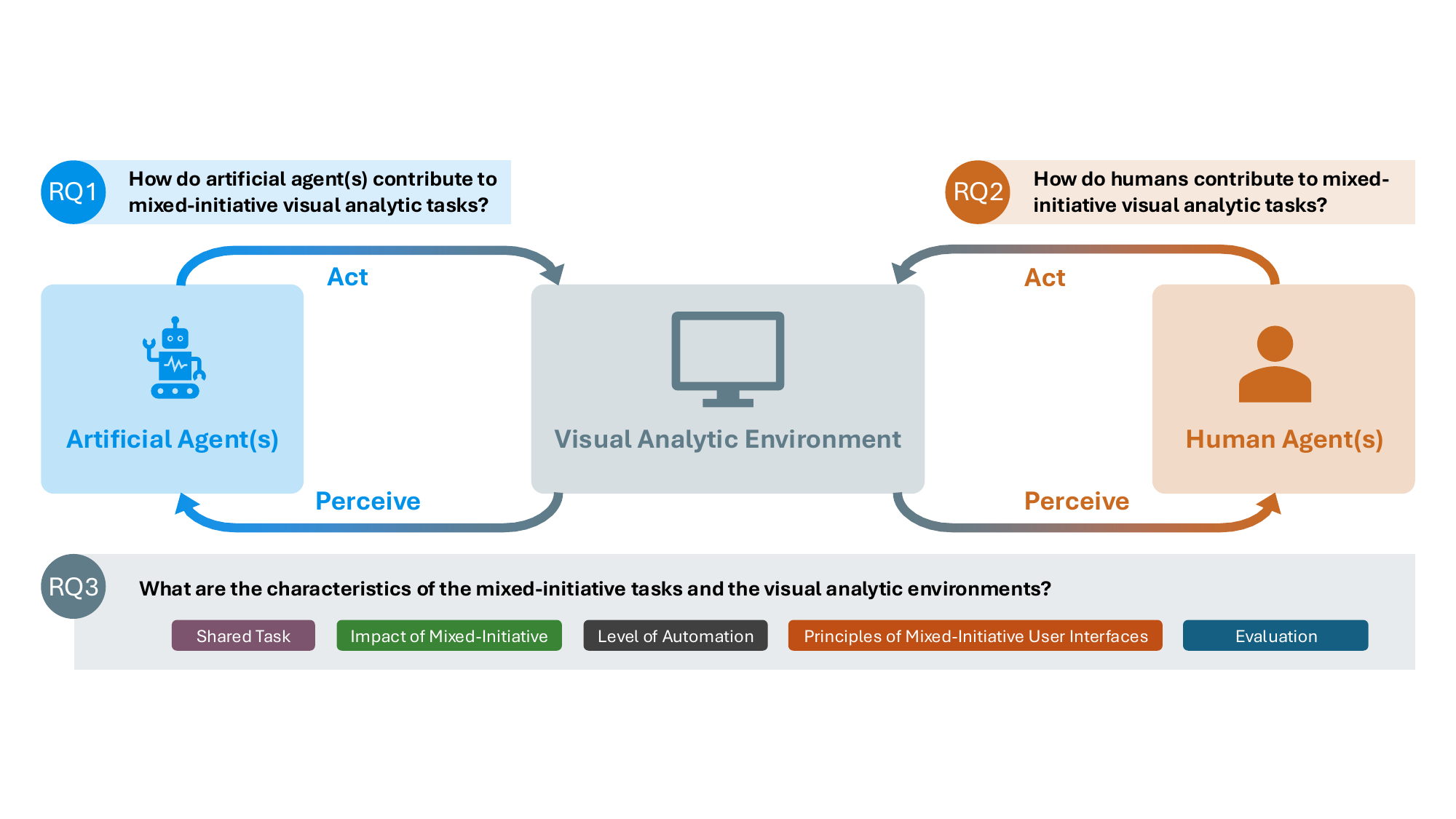}
  \centering
  \caption{The overview of the research questions in our scoping review, where we aim to characterize \emph{mixed-initiative} visual analytic settings. Throughout this work, we adopt an agent-based framework where human and artificial agents perceive their visual analytic environment and act upon it.}
  \label{fig:teaser}
}

\maketitle
\begin{abstract}
Artificial agents are increasingly integrated into data analysis workflows, carrying out tasks that were primarily done by humans. Our research explores how the introduction of automation recalibrates the dynamic between humans and automating technology. To explore this question, we conducted a scoping review encompassing twenty years of mixed-initiative visual analytic systems. To describe and contrast the relationship between humans and automation, we developed an integrated taxonomy to delineate the objectives of these mixed-initiative visual analytics tools, how much automation they support, and the assumed roles of humans. Here, we describe our qualitative approach of integrating existing theoretical frameworks with new codes we developed. Our analysis shows that the visualization research literature lacks consensus on the definition of mixed-initiative systems and explores a limited potential of the collaborative interaction landscape between people and automation. Our research provides a scaffold to advance the discussion of human-AI collaboration during visual data analysis. 
Our integrated taxonomy is available in the form of a web application on \url{https://smonadjemi.github.io/miva}.

\end{abstract}  

\section{Introduction}

Modern artificial intelligence (\acro{AI}) tools, supported by various foundation models, have demonstrated extensive capabilities toward a variety of tasks including data analysis and visualization generation.
For example, Data Analyst by OpenAI accepts data files and natural language prompts as inputs, and it responds with analysis, insights, and visualizations.
Yet, these advances occur against a backdrop of mixed-initiative visual analytic systems that have tackled many of the same problems using different ad hoc approaches over the years.
In this research, we examine the characteristics of mixed-initiative visual analytic systems to delineate their design space and reflect on the potential application, or disruption, that modern \acro{AI} technology introduces. 

Visual analytic (\acro{VA}) tools integrate large data sources, advanced modeling techniques, and inherent human strengths for more effective data analysis \cite{keim2008visual}. 
This integration is facilitated by interactive visual interfaces that empower humans to engage with their data and underlying models seamlessly.
To further enhance the data analysis workflows, visual analytics researchers have designed artificial agents that provide analytic guidance and automate parts of the tasks, leading to \emph{mixed-initiative} visual analytic systems \cite{sperrle2023lotse, ceneda2017characterizing}.  
These artificial agents (often referred to as guidance engines in \acro{VA} literature) are powered by a variety of automation techniques, from simple heuristics, to machine learning methods, and more recently, large language models.
While modern \acro{AI} tools rely on natural language dialogues with their human counterparts, artificial agents designed for visual analytic tools often observe and learn from human interactions with the visual interface. These interaction logs capture the steps taken by analysts and shed light on higher-level aspects of reasoning such as cognitive biases and analytic objectives. 
Therefore, mixed-initiative visual analytic systems span a rich and unique design space on how humans and automating technology interact. Being able to reflect on these design choices and consider their impact is informative as our research community continues to explore to what extent and in what ways automation should be used to support visual data analysis.

Here, we present \textit{a scoping review of mixed-initiative visual analytic systems developed over the past twenty years}. The purpose of a scoping review is to conduct a synthesis of literature and characterize how a set of concepts have been defined, presented, and evolved over time~\cite{munn2018systematic}. By doing so, we describe the design space of mixed-initiative visual analytic systems consistently and discuss navigating this design space in future research that integrates emerging \acro{AI} technology. 
To conduct our scoping review, we undertook two phases of analysis, consisting of constructing, extending, and applying a set of qualitative codes on collected literature. We alternated rounds of top-down (theory-driven) and bottom-up (data-driven via grounded theory) approaches. 
Our top-down approach integrated codes from existing taxonomies, typologies, and theories on mixed-initiative systems (e.g., \cite{horvitz1999principles, domova2023a, parasuraman2000a}) to describe and classify the characteristics of mixed-initiative \acro{VA} systems. Our bottom-up approach involved identifying and extracting descriptive text excerpts from our literature corpus to identify additional characteristics of these systems beyond what existing theory describes. Over these two phases, we analyzed 86 papers. The result is \textit{an integrated taxonomy for describing the design space of mixed-initiative visual analytic systems.}  Our integrated taxonomy considers mixed-initiative systems in terms of 80 defining characteristics spanning 7 broad attributes (Figure~\ref{fig:teaser}). These categories describe the role of humans and automating agents, the types and levels of automation, how these systems are assessed, and their alignment to principles of mixed-initiative interfaces outlined by \citet{horvitz1999principles}.  
Collectively, our research contributions are:
\begin{itemize}[]
    \item A review of mixed-initiative visual analytics research from the past two decades, collected systematically and analyzed using theory-driven and grounded theory qualitative techniques.
    \item An integrated taxonomy to formalize the design space of mixed-initiative visual analytics systems in terms of human agent(s), artificial agent(s), and analytic environments.
    \item Reflection on the characteristics of existing mixed-initiative visual analytic systems and a vision for future work on enabling human-AI collaboration for visual data analysis.    
\end{itemize}

It is pertinent and timely to consider whether the affordances of modern \acro{AI} tools materially advance the design space of mixed-initiative visual analytics, or whether they just offer a, potentially problematic, alternative avenue.
Our research presents a way to surface and organize what has already been done and to reflect on the role of new techniques. As researchers will undoubtedly continue to examine and integrate modern \acro{AI} tools into visual data analysis workflows, our contributions present a way to ground those investigations and contextualize their contributions.

\section{Background}\label{sec:related_work}

\subsection{Mixed-Initiative Systems Broadly}
\label{ss:mi_broadly}

The idea of \emph{mixed-initiative} systems dates back to before the 1970s. Although not always labeled as ``mixed-initiative'', early work was interested in humans and computers jointly controlling physical devices such as undersea robots \cite{sheridan1978human} and aircraft \cite{mcgee1998future}. In early days, the source of computer reasoning was mainly rule-based agents that observed their environment through sensors and took pre-programmed, deterministic actions autonomously given the circumstances. On the human side, there was often a physical interface (e.g., a control board with buttons), where the human operator was afforded an action space to control the remote device or supervise its autonomous actions as needed. Realizing the undesirable consequences of over-relying on automation \cite{wiener1987fallible}, especially in safety-critical missions, Riley proposed the \emph{mixed-initiative model} as a formalization of human-machine interaction in joint decision-making. In this model, Riley envisioned mixed-initiative systems to be comprised of \emph{machines} and \emph{operators} acting upon the \emph{world} in order to accomplish a \emph{shared goal}. Furthermore, a two-dimensional taxonomy was proposed for characterizing automation: level of intelligence (relevant knowledge) and level of autonomy (jurisdiction to act) \cite{riley1989a}. 
This taxonomy was designed to help practitioners decide \emph{when} and \emph{how much} automation is appropriate (a question still relevant today \cite{simmler2021taxonomy}). 

Over time, \acro{AI} agents emerged that could learn from observations and make decisions strategically, without pre-programmed rules being given to them.
These advances paved the way for more adoption of automation across different tasks. \citet{parasuraman2000a} enumerate four types of tasks that may benefit from automation (information acquisition, analysis, decision selection, and action implementation). Each of these tasks may be automated to various levels from low-automation to high-automation, depending on the degree of human direction (e.g., \cite{sheridan1978human}). This taxonomy has been tailored to the visual analytics domain by \citet{domova2023a}. There exist other taxonomies and typologies that aim to characterize the tasks and levels of automation in mixed-initiative systems. See the survey by \citet{vagia2016a} for more details. In this scoping review, we utilize the 10-degree spectrum proposed by \citet{parasuraman2000a} to characterize the \leavevmode{\textbf{\color{levelOfAutoColor}{level of automation}}} in mixed-initiative visual analytics systems. Furthermore, we draw inspiration from the work by \citet{domova2023a} to characterize the \leavevmode{\textbf{\color{sharedGoalsColor}{shared task}}} along the four core components of visual analytics: data, models, visualizations, and decisions/knowledge \cite{keim2008visual}.

\subsection{Automation in Visual Analytic Workflows}

If humans and computers jointly manipulate physical devices such as undersea robots \cite{sheridan1978human}, then why could (or should) they not jointly manipulate visual interfaces and conduct data analysis? 
This was a debate in the 1990s where some researchers such as Pattie Maes advocated for \emph{interface agents} and others such as Ben Shneiderman advocated for \emph{direct manipulation} \cite{shneiderman1997direct}.
Despite some hesitation against integrating artificial agents into user interfaces, researchers and software companies continued pursuing the idea and even included some interface agents in their products (e.g., Microsoft Clippy in 1996).
As we now know, Clippy turned out to be an inconvenience for most users, as it offered help when no help was needed \cite{stratton_introduction_2024}.
To formalize effective integration of artificial agents into user interfaces, \citet{horvitz1999principles} outlined a list of twelve principles for mixed-initiative user interfaces. Adherence to these principles became one way to define a user interface to be \emph{mixed-initiative} (e.g., Podium \cite{wall2018podium}). Here, we characterize existing mixed-initiative visual analytic tools in terms of these \leavevmode{\textbf{\color{miPrinciplesColor}{principles of mixed-initiative user interfaces}}}.

Continuing to investigate how automation fits user workflows, researchers also considered \emph{analytic} tasks.
\citet{bertini2010investigating} offer one of the early discussions on both automatic and visualization-based data analysis techniques, and describe a mixed-initiative knowledge discovery pipeline. For example, consider the task of uncovering the relationship between two numeric attributes in a dataset. On one hand, users may observe a scatter plot of a dataset and arrive at a conclusion about the relationship between two variables. On the other hand, a mathematical model can be fitted to represent the patterns in the data automatically. \citet{bertini2010investigating} present these two scenarios as two extreme ends of knowledge discovery (fully manual and fully automated respectively) and argue that when neither human nor the machine is dominant in this joint cognition, then we have accomplished a  \emph{mixed-initiative} knowledge discovery pipeline. While this view of \emph{mixed-initiative} systems focuses on higher-level cognition, it does not consider the actions taken by agents. In a recent conceptual model, \citet{Satyanarayan2024Intelligence} reflect on the commercialization of \acro{AI} chat-bots powered by large language models and how humans will interact with \acro{AI} as active agents (as opposed to inanimate objects). 
They argue for redefining intelligence as `agency' (the ability to meaningfully act) and focusing on the delegation of constrained agency between human and \acro{AI} agents \cite{Satyanarayan2024Intelligence}. In our scoping review, we are interested in characterizing how existing mixed-initiative visual analytic systems delegate agency to \leavevmode{\textbf{\color{humanAgentColor}{human}}} and \leavevmode{\textbf{\color{aiAgentColor}{artificial}}} agent(s).

Another common way to view mixed-initiative visual analytic systems is through the lens of \emph{guidance}. \citet{ceneda2017characterizing} define analytic guidance as a computer-assisted process to resolve analyst knowledge gaps. 
They offer a framework to characterize the nature of the knowledge gap, the inputs and outputs of the guidance algorithm, and the degree of guidance offered to users.
\citet{collins2018guidance} follow up by outlining six broad goals for adopting mixed-initiative approaches in analytic settings (e.g., to reduce cognitive load, etc.).
To promote wider adoption of analytic guidance in practice, \citet{sperrle2023lotse} propose Lotse, a library for configuring guidance strategies into structured \texttt{yaml} files. 
Lotse aims to make it easier to design, deploy and reuse guidance approaches in visual analytic settings. While the systems we reviewed adopted ad-hoc techniques for guidance, a reusable gallery of analytic guidance agents could result in replicability and benchmark studies. Our scoping review contributes a running list of common guidance mechanisms in existing mixed-initiative visual analytics systems, hence providing the grounds for implementing them as reusable artificial agents using tools such as Lotse.

In the era of greater access to artificial agents, we anticipate that the roles of humans and \acro{AI} will evolve to where either can take fully autonomous actions and collaborate with one another in refining their actions. 
This anticipation has spurred new research, including ours, to understand the types of analytical tasks machines can effectively perform \cite{perez2022a} and develop methods for ensuring humans and artificial agents adapt to one another throughout a session \cite{sperrle2021coadaptive}.
\added{
Recent work by \citet{lin2023promptsexploringdesignspace} demonstrates the importance of co-creation and mixed-initiative interaction on user experience and quality of outcomes.
}
Our scoping review sheds light on how humans and artificial agents share agency in mixed-initiative systems targeted towards visual data analysis.

\section{Problem Definition}\label{sec:problem_definition}

We approach mixed-initiative visual analytics using an agent-based framework \cite{monadjemi_humancomputer_2023} and dissect existing systems into three components shown in Fig.~\ref{fig:teaser}: human agent(s), artificial agent(s), and the visual analytic environment. We seek to characterize each component by answering the following questions:
\begin{enumerate}[label=(RQ\arabic*), leftmargin=*]
    \item How do \leavevmode{\textbf{\color{aiAgentColor}{artificial agent(s)}}} contribute to the analytic tasks?
    \item How do \leavevmode{\textbf{\color{humanAgentColor}{human(s)}}} contribute to the analytic tasks?
    \item What are the characteristics of the mixed-initiative tasks and the visual analytic environments? 
\end{enumerate}

For RQ1 and RQ2, we will leverage the taxonomy \citet{holstein2020a} to characterize \emph{what} kind of contribution \leavevmode{\textbf{\color{humanAgentColor}{human}}} and \leavevmode{\textbf{\color{aiAgentColor}{artificial}}} agents make. Furthermore, we extract \emph{how} these contributions appear in existing systems based on text excerpts from the reviewed literature.
For RQ3, we characterize the visual analytic tasks and environments in terms of five attributes: 
\leavevmode{\textbf{\color{sharedGoalsColor}{shared task}}},
\leavevmode{\textbf{\color{miImpactColor}{impact of a mixed-initiative approach}}},
\leavevmode{\textbf{\color{levelOfAutoColor}{level of automation}}},
\leavevmode{\textbf{\color{miPrinciplesColor}{principles of mixed-initiative user interfaces}}}, and
\leavevmode{\textbf{\color{evaluationColor}{evaluation technique}}}.
Similarly, we leverage terminology from existing theories as well as details extracted from text content of reviewed work for each of these attributes.

\section{Methods}\label{sec:survey_method}

\noindent\textbf{Scoping Review Approach.} In a scoping review, the data of interest is the text content from research papers selected via a systematic process and the goal is to characterize how emergent themes interact with one another. To analyze this form of unstructured data, researchers assign categories and descriptors to blocks of text in a process called \emph{qualitative coding} \cite{lazar2017research}.
There are two common coding approaches, both of which have been leveraged in \acro{HCI} research. 
\emph{Emergent coding} (also known as bottom-up or grounded theory) is the process of reading the documents with an open mind and free of any pre-existing expectations, hence establishing theories that emerge from the data \cite{glaser1999the}. 
Alternatively, \emph{a-priori coding} (also known as top-down or theory-driven) is an approach where existing theories and taxonomies are used to inform the categorization of the text content. In fields where existing work is limited, researchers often opt for emergent coding, which is typically a more difficult process. In fields where there are existing theories and taxonomies that are well-accepted and applicable to the analysis at hand, researchers opt for the \textit{a priori} coding paradigm, which is typically easier. In some instances such as the work by Feng et al.\ \cite{feng2010computer}, researchers leverage a hybrid approach to maximize the usage of established taxonomies, while also uncovering new concepts not captured in the existing literature.
We take a hybrid approach to our scoping review by identifying emergent themes in addition to leveraging the rich body of taxonomies and typologies proposed by the visual analytics research community, as well as existing theories about mixed-initiative systems in the \acro{AI} community. This approach allows us to use existing taxonomies and typologies (as opposed to proposing yet another similar one), while maintaining flexibility over new themes emerging during our analysis.

\begin{figure*}[!t]
    \centering
    \includegraphics[clip, trim=0cm 0.5cm 0cm 0.5cm, width=0.9\textwidth]{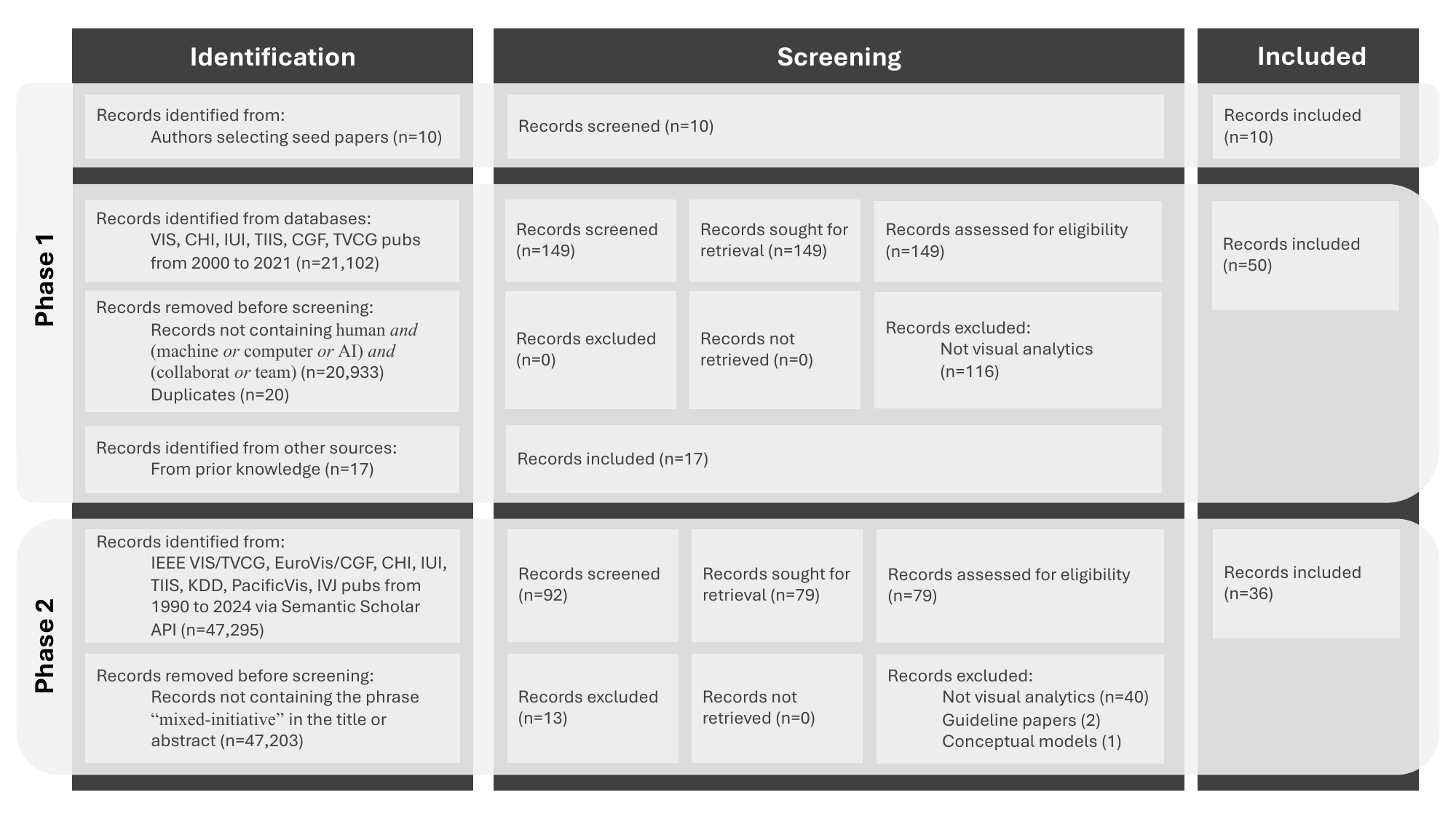}
    \caption{The PRISMA-ScR diagram summarizing our literature review process.}
    \label{fig:prisma}
\end{figure*}

\vspace{1mm}
\noindent\textbf{Multi-phased Execution.} We established a two-phase process for our scoping review (summarized in Fig.\ \ref{fig:prisma}). First (\S\ref{sec:survey_codebook_ideation}), we reviewed 60 papers to help us externalize the defining aspects of \emph{mixed-initiative} visual analytics systems and how they compare to the broader categories such as \emph{human-computer collaboration} for visual data analysis. 
During this phase, we drew parallels between our observations and existing taxonomies concerning automation and mixed-initiative systems. Whenever possible (i.e., without significant loss of meaning), we used the established terminology as codes, rather than creating new ones.
After this phase, we had a robust codebook covering aspects of mixed-initiative \acro{VA} systems which we refer to as an \emph{integrated taxonomy}. We then conducted a review of 36 papers that self-identify as mixed-initiative visual analytics (\S\ref{ss:phase3_methods}) to assess our integrated taxonomy's comprehensiveness and extend it as needed. We now provide additional details for each of these two analysis phases.

\subsection{Phase 1: Develop Codebook}\label{sec:survey_codebook_ideation}
\noindent\textbf{Identify Salient Characteristics.}
We began codebook ideation through a bottom-up approach. Each author selected two papers that they thought would clearly exemplify the area of mixed-initiative visual analytics. 
The resultant set of ten papers spanned the 2011-2023 time frame (see supplemental material).
All authors reviewed this set, taking notes on aspects that stood out as defining characteristics and the design space of \emph{mixed-initiative} visual analytics systems. Over the course of a month, we met to discuss our coding process, justify the rationale behind our choices, clarify definitions, and discuss conflicts.  Through these iterative discussions and refinement of our ideas, this phase solidified our high-level understanding toward a superset of the most salient aspects of mixed-initiative visual analytics systems including: \textbf{shared tasks} (e.g., topic modeling), \textbf{user groups} (e.g., domain experts), \textbf{labor division} between agents, and \textbf{automating methodology} (e.g., machine learning, domain-specific language, etc.).

\noindent\textbf{Expand Codebook with Details.}
Next, we reviewed a sample of 50 papers related to \emph{human-computer collaboration} for visual data analysis to further develop our codebook and understand the nuances of the terms \emph{mixed-initiative}, \emph{human-\acro{AI} collaboration}, and \emph{human-computer collaboration} that are sometimes used interchangeably in the literature.
These papers were collected as part of prior work \cite{monadjemi_humancomputer_2023} \added{which are related to human-AI teaming.} \deleted{by searching the proceedings of major human-computer interaction and visualization venues and identifying papers that contain the terms related to human-machine/\acro{AI}/computer collaboration or teaming.}
\added{The collection consisted of 17 papers selected from the authors' prior knowledge and 33 papers retrieved from relevant proceedings. To retrieve the papers, a keyword search was applied to publications from major human-computer interaction and visualization venues (VIS, CHI, IUI, TIIS, CGF, TVCG) between 2000 and 2021. The keywords used were ``human'' (for the human aspect), ``machine'' or ``computer'' or ``AI'' (for the AI aspect), and ``collaborat'' or ``team'' (for the collaboration aspect). To be included, a paper must contain at least one keyword from each of the three aspects, e.g., mentioning human, computer, and collaboration at the same time. This resulted in 149 papers for screening. These papers were then retrieved and reviewed, out of which 116 were excluded as they were not related to visual analytics.}
The collection process is shown in the supplemental material.
Each paper was reviewed by two coders (authors of this work), where they took notes on the details of how each salient characteristic was present in systems under review. 
Then, the coders met to discuss their codes and resolve conflicts into one of the three outcomes:
(1) one choice of codes was selected over the other; (2) the two sets of codes were merged; (3) a new set of codes was chosen.  There were no instances where conflicts could not be resolved. We did not track the prevalence of each outcome.

\vspace{1mm}
\noindent\textbf{Incorporate Existing Taxonomies.}
We noticed that many of the emergent themes that we identified also had some coverage with existing taxonomies and conceptual frameworks.
As a group, we mapped elements of our codebook onto taxonomies and conceptual frameworks. To conduct this mapping we explored the literature for prior work pertaining to the themes we identified in the first coding round. For example, to delineate the \leavevmode{\textbf{\color{levelOfAutoColor}{level of automation}}}, we opted to employ a widely used taxonomy by  Parasuraman et al.\ \cite{parasuraman2000a}. Their work describes ten levels of automation that at the extremes have a human or AI agent perform all the tasks independently. 
One consideration we had to make while selecting taxonomies was whether the level of detail provided was adequate for delineating the design space of mixed-initiative systems. For example, there are task and interaction taxonomies (e.g., \cite{perez2022a}) that we deemed to be too abstract to meaningfully capture the interplay between humans and \acro{AI} in mixed-initiative systems. As such, we did not incorporate them into our codebook.  We continued this process for all the themes, identifying relevant taxonomies and integrating them into our codebook.  The result was an \emph{integrated taxonomy} that combined our bottom-up analysis with established frameworks to characterize:

\begin{itemize}[noitemsep]
    \item the contribution of \leavevmode{\textbf{\color{humanAgentColor}{human agent(s)}}} to the shared task \cite{holstein2020a},
    \item contribution of \leavevmode{\textbf{\color{aiAgentColor}{artificial agent(s)}}} to the task \cite{perez2022a, holstein2020a},
    \item the \leavevmode{\textbf{\color{sharedGoalsColor}{shared task}}} between humans and automation \cite{keim2008visual},
    \item the \leavevmode{\textbf{\color{miImpactColor}{impact}}} of adopting a mixed-initiative approach,
    \item the \leavevmode{\textbf{\color{levelOfAutoColor}{automation level}}} denoting the division of agency \cite{parasuraman2000a},
    \item the observed \leavevmode{\textbf{\color{miPrinciplesColor}{principles of mixed-initiative interfaces}}} \cite{horvitz1999principles}.
    \item the \leavevmode{\textbf{\color{evaluationColor}{evaluation techniques}}} used in the study.
\end{itemize}

\subsection{Phase 2: Apply Codebook}
\label{ss:phase3_methods}

\textbf{Paper Selection.}
Through a systematic process, we extracted a set of papers on visual analytics in the past two decades that \emph{self-identify} as mixed-initiative (i.e., through keywords in the title or abstract). There may be mixed-initiative \acro{VA} systems that do not use this specific terminology, possibly stemming from the ambiguity of how mixed-initiative systems should be defined (an issue we seek to address in our work). However, conducting an exhaustive search to account for all cases is not feasible.
\added{Therefore, we decided to focus on visual analytics papers that \textbf{self-identify as mixed-initiative}. Our criteria include: (1) the paper should explicitly claim that its contribution relates to ``mixed-initiative''; (2) the paper should contribute to visual analytics (user interfaces that do not involve data analysis are excluded); (3) the paper should present a method or system (theoretical papers or empirical studies are excluded).}
We first used the Semantic Scholar bulk search \acro{API}\footnote{Semantic Scholar API, accessed Aug 2024. Available at: \url{https://www.semanticscholar.org/product/api}.} to extract all the papers from visualization-related venues (IEEE VIS/TVCG, EuroVis/CGF, ACM CHI, ACM IUI/TIIS, ACM KDD, PacificVis, IVJ) since 1990. \deleted{Then, we wrote a Python script to filter papers that self-identify as mixed-initiative.} \added{Then, we applied keyword search to the collection,} filtering for papers that included the term \textit{mixed initiative} (including different spellings or hyphenations) in the title or abstract. This resulted in 92 records. A total of 13 of the 92 records were excluded by manual review of the title and abstract, which were not academic papers, irrelevant to visualization research, or were not mixed-initiative systems (e.g., mentioning mixed-initiative as an example). We went through the remaining 79 papers and excluded those that were not related to visual analytics (e.g., papers that work on user interface but do not involve visual analytics) or were theoretical papers (i.e., presenting guidelines or conceptual models). This process left us with 36 mixed-initiative visual analytics papers to code. We include the \acro{PRISMA} diagram of this process in the supplemental material.

\vspace{1mm}
\noindent\textbf{Coding Process}
Each paper was assigned to two coders (authors of this paper) for annotation, such that each coder had a total of 16 papers. 
We used the codebook from Phase 1 to characterize (annotate) each mixed-initiative system along the seven high-level attributes (Fig.~\ref{fig:teaser}). In the majority of cases, these characterizations were drawn directly from the codebook. In a small number of cases, a new code was added. 
This characterization process involved highlighting text excerpts from the papers and annotating them. This process achieved two objectives: (1) to recall our reasoning while resolving coding conflicts, and (2) to qualitatively characterize how each attribute has been implemented in the existing work. 
Once we completed the coding, coders met in pairs to discuss and resolve their coding conflicts. Similar to Phase 1, two coders met and looked at their codes item by item. They discussed the differences and their thought process behind selections. The outcomes of the discussions were: (1) picking one coder's choice over the other, or (2) an amalgamation of codes that captures the coded phenomena better than either two individual codes, or (3) a new code. 
To quantify coders' agreement prior to discussions, we computed the inter-coder reliability (ICR) as:
\begin{equation}
    \acro{ICR} = \frac{\mid \mathbf{S_1}  \cap \mathbf{S_2} \mid}{\mid \mathbf{S_1}  \cup \mathbf{S_2} \mid},
\end{equation}
where $\mathbf{S_1}$ is the set of tags assigned to a paper by Coder 1 and $\mathbf{S_2}$ is the set of tags assigned to the same paper by Coder 2.
When comparing \acro{ICR} values across attributes, we observed the range of $0.45$ (least agreement for Principles of Mixed-Initiative User Interfaces) to $0.74$ (most agreement for Artificial Agents' Means of Contribution). We include more details of the \acro{ICR} values in the supplemental material.
Upon reflecting on the \acro{ICR} values, we observed that factors such as the presentation of related information in the reviewed papers impacted the coders' abilities to identify the information. 
Through discussion over the course of several months, we resolved these conflicts by referring back to the reviewed manuscripts to exchange evidence on how we came to our coding decisions and considering more information such as the system implementation details.
The results we present in the remainder of this paper are based on the data after coders' consensus.

\section{Results: An Integrated Taxonomy for Mixed-Initiative VA}\label{sec:survey_results}

\subsection{Human and AI Contributions}
\label{ss:results-human-contribution}

The design of mixed-initiative systems expresses a view on \textit{how}, through interactions with the system, people and automation bring their respective strengths to visual analytic workflows. In this section, we describe how humans and automation initiate interactions and contribute to analytic tasks in the reviewed manuscripts.
Drawing on human-AI collaboration taxonomy by \cite{holstein2020a}, we describe four types of contributions:  goal augmentation, action augmentation, cognitive augmentation\footnote{\label{cognitive_footnote}~\citet{holstein2020a} use the term \emph{perceptual augmentation} to describe this type of contribution. However, because the term “perception” in the visualization literature typically refers to visual sensing, we adopt the term \emph{cognitive augmentation} instead. This better emphasizes the cognitive dimension of an agent's contribution in mixed-initiative settings, which more closely aligns with both the taxonomy and our intended use of the term in this work.}, and decision augmentation. 
In mixed-initiative settings, either the {\textbf{\color{humanAgentColor}{human}}} or {\textbf{\color{aiAgentColor}{artificial agents}}} can make contributions to the shared task.

\textit{\textbf{Goal augmentation is the process of an agent enhancing another agent's understanding of an analytic goal through clarification and added information.}}  We observed {\textbf{\color{humanAgentColor}{analysts}}} communicating information about their desirable goal states by providing examples, or by providing constraints. For example, MultiVision is a mixed-initiative system for designing dashboards where \emph{``the optimizer takes optional {\textbf{\color{humanAgentColor}{user}}} specification as constraints to generate conditional recommendations''} \cite{wu2022multivision}. As another example, Intentable is a mixed-initiative tool for generating captions, where \emph{``the {\textbf{\color{humanAgentColor}{user}}} can customize the number of caption sentences, the order between sentences, and the content of each sentence''}, hence specifying constraints for generated captions. Another way that {\textbf{\color{humanAgentColor}{humans}}} provide information is by selecting seed examples. \cite{cui2022a} present a mixed-initiative system where the {\textbf{\color{humanAgentColor}{human}}} is responsible for making subjective choices on which visualization design is appropriate for a task at hand and providing the system with seed examples. The artificial agent, in return takes further actions that are aligned with the goal (inferred from the seed examples).
In visual analytic workflows, humans are assumed to be in control of the analysis and setting their goals. Hence, we did not observe any instances where the artificial agent sets or augments the goal explicitly. Rather, goal augmentation by the artificial agent may occur implicitly through recommendations, discussed next.

\textit{\textbf{Action augmentation refers to modifying the set of actions under consideration by either the human or the artificial agent.}}
Here, we focus on the interactions between the human and artificial agents, as opposed to their independent actions.
We noticed two prominent ways {\textbf{\color{humanAgentColor}{human agents}}} initiate change in the action space: by prompting another agent to recommend actions, or by manipulating the recommended actions and hence creating a new (better) action. 
In some systems such as PlotThread~\cite{tang2021plotthread} and MEDLEY~\cite{pandey2023medley}, {\textbf{\color{humanAgentColor}{users}}} can explicitly ask an artificial agent for recommendations to move their analysis forward. In a use case for PlotThread, \emph{``[{\textbf{\color{humanAgentColor}{the user}}}] thinks the appearance is not legible or aesthetically appealing enough, so he seeks for AI assistance by triggering the AI creator''}~\cite{tang2021plotthread}. This results in a set of \emph{``{\textbf{\color{aiAgentColor}{AI}}}-generated storylines [being] displayed in a list''} as recommendations for the user to consider.
Sometimes, the user can further refine these recommendations by making manual edits.
For example, Causemos \cite{husain2021a} is a mixed-initiative system for causal modeling that \emph{``provides [{\textbf{\color{humanAgentColor}{users}}}] bulk editing actions to correct concept assignments, change polarity, or discard statements altogether''} while constructing causal maps.
In some mixed-initiative systems, the {\textbf{\color{aiAgentColor}{artificial agent}}} may make recommendations without explicit requests from the user.
For example, the {\textbf{\color{aiAgentColor}{artificial agent}}} in \mbox{Voyager~2~\cite{wongsuphasawat2017voyager2}} \emph{``automatically recommends charts based on the current user-specified focus view, promoting discovery of relevant data fields and alternative ways to summarize or encode the data.''}
In other cases, the {\textbf{\color{aiAgentColor}{artificial agent}}} may generate the action space and rely on the analyst to choose an action. For example in \mbox{MultiVision~\cite{wu2022multivision}}, \emph{``given an input data table, the automation module leverages deep learning models to predict an assessment score for all possible charts''}, hence generating the action space and some predicted measure of quality for analysts' consideration.

In contrast to action augmentation, which aims to curate a set of possible actions,  \textit{\textbf{decision augmentation is about selecting an action}}. We observed three primary ways {\textbf{\color{humanAgentColor}{users}}} augment decisions in our reviewed mixed-initiative settings. In Podium~\cite{wall2018podium}, a mixed-initiative system for ranking data points and setting weights for feature importance, {\textbf{\color{humanAgentColor}{users}}} can directly assign values to the parameters, hence directly making decisions in the context of that analytic task. In systems like Kori~\cite{latif2022kori} and the topic modeling system proposed by \citet{sperrle2021learning}, the user is presented with an action by the {\textbf{\color{aiAgentColor}{artificial agent}}} and that action is not taken until the {\textbf{\color{humanAgentColor}{user}}} accepts the recommendation. Finally, in systems like Active Data Biology \cite{dasgupta2017familiarity} and the system proposed for extracting data from images by \citet{jung2021mixed}, the {\textbf{\color{aiAgentColor}{artificial agent}}} takes autonomous actions and the {\textbf{\color{humanAgentColor}{user}}} is responsible for either approving the inferred results before continuing with the analysis or undoing the actions taken by the artificial agent.
In our review, we observed {\textbf{\color{aiAgentColor}{artificial agents}}} who make autonomous decisions to update the visual layout or classify topics.
For example, C8~\cite{zhao2017annotation} is a mixed-initiative system for analyzing and annotating graph datasets where \emph{``all nodes are positioned automatically based on their similarities, and a user could choose to manually override the position of specific nodes that are then integrated into the automatic layout.''} This is an example where an {\textbf{\color{aiAgentColor}{artificial agent}}} decides about the visual layout of the data.

Finally, \textit{\textbf{cognitive augmentation\footnotemark[2] refers to cases where an agent contributes with their (sometimes subjective) perspective towards an action or piece of information}}. For {\textbf{\color{humanAgentColor}{human agents}}}, this process may involve inspecting alternatives and determining importance given their situational understanding. For example, FAIRVIS~\cite{cabrera2019fairvis} is a mixed-initiative system in which {\textbf{\color{humanAgentColor}{users}}} perceive the model outputs across subpopulations and determine if there is an opportunity for mitigating model bias for some sub-populations. Similarly, IsoTrotter~\cite{palenik2021isotrotter} is a mixed-initiative system for exploring model parameter space, where {\textbf{\color{humanAgentColor}{users}}} contribute  by maintaining an \emph{``intuition for the functioning of the model and the expected behaviour''} given various inputs. In this setting, {\textbf{\color{humanAgentColor}{users}}} are assumed to be domain experts whose \emph{``ability [...] to make predictions about the model outcome is essential for the model validation and optimization.''}
We also observed {\textbf{\color{aiAgentColor}{artificial agents}}} making cognitive contributions to the task at hand. This contribution often appeared as summarizing text, predicting outcomes, and extracting structured data from unstructured sources.
For example, in TimeFork~\cite{badam2014timefork}, the \acro{AI} agent contributes by predicting future values of time-series data. 

\subsection{Shared Analytic Task}
In mixed-initiative systems, {\textbf{\color{humanAgentColor}{humans}}} and {\textbf{\color{aiAgentColor}{artificial agents}}} make decisions and take actions to contribute to a {\textbf{\color{sharedGoalsColor}{shared~task}}}.
Understanding these shared tasks and analytic goals is critical to delineating the roles and interactions between the agents. This is synonymous with \emph{defining the target of analysis}, the first step identified by \cite{perez2022a} for designing mixed-initiative visual analytic systems.
To frame these shared analytic tasks, we draw inspiration from established models of visual analytics (e.g., \cite{perez2022a,keim2008visual}), which emphasizes the interplay between four major components: data, models, visualizations, and knowledge/decisions. 

\sharedtask{\textit{\textbf{Data-centric tasks focus on enhancing the exploration, cleaning, or transformation of data to support downstream analysis}.}} For example, we observed mixed-initiative systems that assist in \emph{summarizing data}~\cite{zhang2023concepteva}, \emph{discovering relevant data}~\cite{dasgupta2017familiarity}, \emph{extracting data from images}~\cite{jung2021mixed}, \emph{annotating data}~\cite{dinakar2015mixed}, and \emph{ranking data points}~\cite{wall2018podium}.

\sharedtask{\textit{\textbf{Model-centric tasks emphasize building, refining, or validating models.} }}Human and \acro{AI} agents work together to adjust model parameters, evaluate model outputs, or explore alternative hypotheses. Some examples from the reviewed literature include \emph{refining models}~\cite{palenik2021isotrotter}, \emph{inspecting models}~\cite{cabrera2019fairvis}, \emph{analyzing clusters}~\cite{dasgupta2017familiarity}, and \emph{understanding models}~\cite{zhang2023interpreting}.

\sharedtask{\textit{\textbf{Visualization-centric tasks pertain to designing, re-using, or adapting specifications that result in visual representation of the data.}}} For example, Mystique~\cite{chen2024mystique} is a mixed-initiative system that \emph{``extract[s] the axes and legend, and deconstruct[s] a chart's layout into four semantic components''} for easier reuse of info-graphics. As another example, \emph{``VisExemplar is a mixed-initiative data exploration prototype that allows users to explore their data using Visualization by Demonstration''}~\cite{saket2017visualization}. Similarly, Voyager~2~\cite{wongsuphasawat2017voyager2} is a \emph{``mixed-initiative tool that blends manual and automatic chart specification in a unified system''}.

\sharedtask{\textit{\textbf{Knowledge- and decision-centric tasks encompass the broader objective of deriving actionable insights or making informed decisions.}}} For instance, we observed mixed-initiative systems that assist analysts in \emph{discovering evidence}~\cite{chau2011apolochi}, \emph{navigating a design space}~\cite{pu2002design}, \emph{generating insights}~\cite{lin2024inksight}, externalizing knowledge~\cite{cook2015mixed}, and \emph{forecasting outcomes}~\cite{badam2014timefork}.

\subsection{Impact of Mixed-Initiative Visual Analytics}
\label{ss:results-motivations}

So far, we have established that mixed-initiative systems involve {\textbf{\color{humanAgentColor}{humans}}} and {\textbf{\color{aiAgentColor}{artificial agents}}} making decisions and taking actions to contribute to a {\textbf{\color{sharedGoalsColor}{shared~task}}}.
Here we describe our observations pertaining to the {\textbf{\color{miImpactColor}{impact}}} of adopting a mixed-initiative approach for visual data analysis. The (expected or observed) impacts shed light on the incentives, motivations, or justification for opting for mixed-initiative systems over either extremes of humans or automation completing the task alone. 
We identified five key drivers for pursuing mixed-initiative solutions: speed, accuracy, accessibility/convenience, alignment with humans, and domain knowledge. 

\impact{\textit{\textbf{Speed is the most common motivation for choosing a mixed-initiative approach for visual data analysis.}}} The most common rationale is that the task would otherwise be too tedious for a human to do alone. 
For example, \citet{latif2022kori} argue that \emph{``[a]n automatic reference detection intends to speed up the composition process''} while synthesizing text and visualizations in a manuscript.
\citet{husain2021a} argue that \emph{``[non mixed-initiative] systems are limited with respect to the speed with which users can build their model''}.
Many authors argue that the time analysts gain by using mixed-initiative systems can be instead allocated to tasks that are more unique to humans such as \emph{``pursue scientific questions''} \cite{dasgupta2017familiarity} and \emph{``investigate the why aspect of their findings''} \cite{dasgupta2017familiarity}.
We note that none of the papers in our pool were motivated by a mixed-initiative approach being faster than automation alone. We will further discuss this observation in \S\ref{sec:discussion}.

\impact{\textit{\textbf{Mixed-initiative systems improve accuracy in comparison to either a human or an artificial agent performing the task alone.}}} We observed this to be a bidirectional relationship where (1) automation is deployed to improve the accuracy of a human conducting the task alone, or (2) the human is integrated in the loop to improve the accuracy of automation performing the task alone.
Let us reconsider Mystique~\cite{chen2024mystique}, a mixed-initiative system for reusing infographic charts. The motivation for this system being mixed-initiative is that \emph{``heuristics don’t guarantee 100\% accuracy, and thus [the authors] build a user interface in Mystique to allow fixing potential errors through simple interactions''}.
In other cases such as LEVA~\cite{zhao2024leva}, the mixed-initiative system was envisioned to lead to more accurate results with help from an artificial agent.

\impact{\textit{\textbf{Mixed-initiative systems also enhance accessibility and convenience.}}} They do so by lowering the cognitive load for users, making complex tasks more manageable, and improving the interpretability of \acro{AI} models. For example, Panwar et al.\ \cite{panwar2018detecting} argue that high cognitive load can increase \emph{``the likelihood of disengagement or making mistakes''}. These systems are particularly beneficial for non-technical users, offering intuitive interfaces and control mechanisms that are more familiar and user-friendly than fully manual or fully automated approaches respectively. 
For example, \cite{wongsuphasawat2017voyager2} propose Voyager~2, a mixed-initiative system for creating visualization dashboards. This system is motivated by the hypothesis that \emph{``providing complete view specifications can be tedious, and require domain familiarity as well as design and analysis expertise''}. Hence, Voyager~2 makes it more convenient and accessible for users to create charts by only deciding on partial specifications.

Finally, mixed-initiative systems, compared to fully autonomous systems, \impact{\textit{\textbf{provide affordances for better alignment with humans' preferences and objectives.}}} For example, \emph{``there often exist multiple, equally correct solutions that fulfill different user preference profiles''} \cite{sperrle2021learning} when curating topic models. Mixed-initiative systems allow for such subjective preferences to be incorporated into models.  \impact{\textit{\textbf{For autonomous systems, a benefit is that they can obtain and integrate domain expertise in real time}}}, ensuring the outputs are relevant and contextually appropriate for the user and task at hand.  For example, an analyst may leverage their objective domain knowledge to \emph{``generate and investigate known subgroups''} while inspecting model performance~\cite{cabrera2019fairvis}.

\subsection{Level of Automation}

We have characterized mixed-initiative systems to involve {\textbf{\color{humanAgentColor}{humans}}} and {\textbf{\color{aiAgentColor}{artificial agents}}} making decisions and taking actions to contribute to a {\textbf{\color{sharedGoalsColor}{shared~task}}} and having an {\textbf{\color{miImpactColor}{impact}}} on the workflow compared to non-mixed-initiative approaches.
An important design choice that we observed in our literature review was the degree to which humans and automation share agency. We refer to this design choice as {\textbf{\color{levelOfAutoColor}{level of automation}}}.
We followed the classification scheme proposed by \citet{parasuraman2000a}, which categorizes the level of automation into ten levels ranging from the computer not providing any assistance (Level 1) to the computer deciding everything and acting autonomously (Level 10).  In our review, we observed only five levels of automation, discussed below.

The earliest mixed-initiative \acro{VA} system in our review is designed to assist in constraint satisfaction tasks such as travel planning and flight scheduling \cite{pu2002design}. This system \autol{\textit{\textbf{offers all available search algorithms}}} as well as \autol{\textit{\textbf{the entire search space}}} while identifying solutions. Given the computational complexity of such search problems and the possibility of the computer not being able to identify a satisfactory solution, the human agent is responsible for observing the search progress and selecting configurations or algorithms that might be put the system on a more promising path to the solution.

More than half of the mixed-initiative \acro{VA} systems we reviewed (20 of 36 papers) offer a slightly higher level of automation where \autol{\textit{\textbf{the artificial agent offers a few options}}} as opposed to the entire set of possibilities. Consider Dupo~\cite{kim2024dupo}, a mixed-initiative tool for creating responsive visualizations. In this system, the artificial agent \emph{``suggests design alternatives''} to \emph{``enable fine-grained refinement of preferred recommendations''}~\cite{kim2024dupo}. Furthermore, this system has a simplified module that \emph{``recommends a single responsive transformation as a possible next step''}, serving as an example of a system operating at different levels of automation by also including a module that \autol{\textit{\textbf{suggests only one option}}}. While we focus our examples on Dupo for creating responsive visualizations, we note that we observed these levels of automation applied to other analytic tasks including model selection and tuning~\cite{elassady2019visual}, visualization specification~\cite{saket2017visualization, wongsuphasawat2017voyager2}, dashboard composition~\cite{wu2022multivision, pandey2023medley, tang2021plotthread, zhang2023interpreting}, and data exploration~\cite{chau2011apolochi, dasgupta2017familiarity, husain2021a, cabrera2019fairvis}.

The levels of automation above narrow down the action space and present the user with a strategically curated set. At higher levels of automation, agents can select and execute an action autonomously. Such higher levels of automation are less explored in the mixed-initiative \acro{VA} literature. Consider the system proposed by \citet{sperrle2021learning} for curating topic models. The action space in this system involves merging or splitting sets of documents to curate cohesive topics. The artificial agent in this system proposes merge/split actions and allows the user some time to accept or decline the recommendation. If the user clicks on the accept button, the system automatically performs the proposed action and updates the models. This serves as an example of a mixed-initiative \acro{VA} system that \autol{\textit{\textbf{acts autonomously upon human approval}}}.

The highest level of automation observed in our search is when \autol{\textbf{\textit{the system acts autonomously and informs the human}}}. For example, consider Voyager~\cite{wongsuphasawat2016voyager} which automatically determines the full specifications of a visualization and presents it to the user, hence informing the user about the decision.  
We emphasize that being \emph{mixed-initiative} does not necessarily require operating at a high level of automation. Rather, it is about choosing the right level of automation for the task, user group, and circumstances. We found it interesting that while Voyager~\cite{wongsuphasawat2016voyager} adopted a high level of automation that fully specified the design of visualizations, the subsequent work, Voyager~2~\cite{wongsuphasawat2017voyager2}, adopted a lower level of automation that only determined partial visualization specifications and allowed users to manually edit the remaining specifications. This reduction in the level of automation allows human agents to express their preferences in light of the analytic task at hand.

\subsection{Evaluation of the Mixed-Initiative Setting}

Many of the systems we reviewed were motivated by the hypothesis that mixed-initiative interaction outperforms either humans or automation performing the task alone, as characterized in \S\ref{ss:results-motivations}. Hence, we paid close attention to the evaluation sections of the reviewed work to extract how this superiority of mixed-initiative visual analytic systems is demonstrated with empirical evidence.

The most common evaluation method in our review (17 of 36 papers) is to conduct \eval{\textit{\textbf{interview studies}}} with potential end users to elicit their feedback on systems. While this form of evaluation provides qualitative insights into how users perceive mixed-initiative systems, it does not provide us with quantifiable evidence for the superiority of mixed-initiative visual analytic systems. Other common evaluation themes we observed were to conduct \eval{\textbf{\textit{experiments involving interaction logs}}}, \eval{\textit{\textbf{Likert scale feedback}}}, and \eval{\textit{\textbf{algorithm evaluation}}}. For example, the artificial agents powering MultiVision~\cite{wu2022multivision} and Kori~\cite{latif2022kori} were evaluated against other algorithmic techniques and the systems themselves were evaluated through single-condition user studies. While this evaluation approach validates the underlying artificial agents and uncovers user sentiment towards systems, they do not provide conclusive evidence on the superiority of mixed-initiative interaction.
Lastly, we observed a significant number of systems being accompanied by \eval{\textit{\textbf{case studies}}}, a hypothetical narrative demonstrating how a system may be used and benefit users. 
We note that \eval{\textit{\textbf{only 8 out of 36 papers in our reviewed pool of papers included empirical comparison of their proposed mixed-initiative setting to either a human or an autonomous agent performing the task alone}}}. For example, both Voyager~\cite{wongsuphasawat2016voyager} and Voyager~2~\cite{wongsuphasawat2017voyager2} were compared against PoleStar, a fully manual visualization tool.

\subsection{Observed Principles of Mixed-Initiative User Interfaces}
\label{ss:results_principles}

\citet{horvitz1999principles} published one of the earliest frameworks of applying mixed-initiative concepts to user interfaces. Expressed through a set of principles, some researchers consider adherence to these principles to be the defining factor for mixed-initiative visual analytic systems. Hence, we attempted to characterize \emph{if} and \emph{how} the reviewed systems adhere to these principles.
Our results show that some of these principles have enjoyed more adoption than others due to factors such as technical bottlenecks. Nonetheless, these principles were forward-looking at the time they were published, and advances in AI will facilitate their wider adoption.

Mixed-initiative visual analytic systems most commonly offer \principle{\textit{\textbf{efficient refinement of results}}} through ensuring users do not feel \emph{``restricted to only what is suggested by the system''}~\cite{latif2022kori}, allowing them to \emph{``fix [...] mistakes''} made by the autonomous agents~\cite{chen2024mystique}, and \emph{``fine-tuning of visual encoding mappings''}~\cite{wongsuphasawat2017voyager2}.
Another commonly observed principle is to \principle{\textit{\textbf{add significant value through automation}}}. Many of these values manifested in a manner consistent with observed and expected impacts of pursuing a mixed-initiative approach as discussed in \S\ref{ss:results-motivations}. For example, authors highlighted that their mixed-initiative tools \emph{``lower the authoring barrier for average users''} \cite{cui2022a}, encourage \emph{``broader exploration, tool learning, and both serendipitous and controlled discovery''} \cite{wongsuphasawat2017voyager2}, and help users \emph{``sacrifice short-term rewards to achieve long-term planning''} \cite{tang2021plotthread}.

An area of interest in the visualization community is \emph{analytic provenance} which aims to extract information about users and tasks by observing low-level interactions with the interface~\cite{xu2020survey, ragan2015characterizing}.
We identified mixed-initiative visual analytic systems that exhibit principles related to analytic provenance. For example, Podium~\cite{wall2018podium} is a mixed-initiative system that \principle{\textit{\textbf{maintains working memory}}} of \emph{``state information including attribute weights, ranking, and marked rows [...] each time the Rank button is pressed''}. Updates to the working memory then trigger \emph{``the system to derive a new set of attribute weights based on the user’s interactions with the rows in the table,''} hence \principle{\textit{\textbf{continuing to learn by observing}}}. In the system proposed by \citet{sperrle2021learning}, \emph{``each [artificial] agent strives to deliver only suggestions that would be accepted by the user while not making suggestions that would be rejected''}, hence \principle{\textbf{\textit{considering user attention and goal uncertainty at the time of service}}} by observing user interactions.

We observed that a narrow subset of Horvitz's principles have been adopted in mixed-initiative visual analytic systems. As we delegate more control to artificial agents (i.e., higher levels of automation), more of these principles will become relevant. With the adoption of mixed-initiative systems by domain experts, making bad or suboptimal decisions could be costly and jeopardize user trust in the system. Hence, implementing safeguards that \principle{\textbf{\textit{minimize poor action/timing}}} would be necessary.  While operating at the higher levels of automation, particularly in safety-critical domains, enabling users to assert \principle{\textit{\textbf{direct invocation/termination}}} over the artificial could further build trust in the system and prevent undesirable outcomes\footnote{This statement is human-centric. Should an artificial agent also be able to terminate human operations if they are expected to lead to bad outcomes?}. Finally, with large language models being widely accessible, we expect to see more systems fostering \principle{\textbf{\textit{dialog for uncertainty resolution}}} in the future, expanding the human-\acro{AI} interaction to go beyond low-level interactions with the visualization.

\section{Discussion}\label{sec:discussion}

\deleted{Mixed-initiative visual analytic systems pre-date the advent of recent \acro{AI} technology. In this scoping review, we sought to define key properties of these systems, reified in our integrated taxonomy.  Our analysis reveals that mixed-initiative visual analytic systems can vary significantly, from their goals, to specific design choices about their levels of automation and the different contributions of humans and artificial agents to the visual analysis processes.}
\added{
Our findings address the research questions as follows. RQ1 examined artificial agent contributions, which are predominantly recommendation-based. RQ2 focused on human contributions, which primarily involve constraint specification and refinement of system outputs. RQ3 characterized task and environmental properties, revealing a concentration in low levels of automation, limited evaluation baselines, and a focus on speed and accuracy gains from mixed-initiative interaction.
In this section, we reflect on our findings and discuss the limitations and opportunities for future work}.

\subsection{Most Commonly Explored Design Choices}

Our scoping review reveals the most salient design choices for mixed-initiative visual analytics systems. We observe that mixed-initiative visual analytics tools most commonly help with 
\leavevmode{\textbf{\color{sharedGoalsColor}{refining models}}} \added{(6 of 36)}
and 
\leavevmode{\textbf{\color{sharedGoalsColor}{making visualizations}}} \added{(7 of 36)}. 
These systems are motivated by being 
\leavevmode{\textbf{\color{miImpactColor}{faster than humans alone}}} \added{(16 of 36)} 
and 
\leavevmode{\textbf{\color{miImpactColor}{more accurate than \acro{AI} agents alone}}} \added{(10 of 36)}. 
Through mixed-initiative interactions, artificial agents contribute to the task by 
\leavevmode{\textbf{\color{aiAgentColor}{augmenting the analyst's action space through recommendations}}} \added{(19 of 36)}, and human agents contribute to the task by 
\leavevmode{\textbf{\color{humanAgentColor}{specifying constraints for acceptable solutions}}} \added{(18 of 36)}
and
\leavevmode{\textbf{\color{humanAgentColor}{manipulating the recommendations offered by the artificial agent}}} \added{(14 of 36)}.
A disproportionate number of coded systems adopt a lower level of automation where
\leavevmode{\textbf{\color{levelOfAutoColor}{the computer suggests a few options}}} \added{(18 of 36)}, hence reducing a large action space. \deleted{We believe the adoption of lower levels of automation is due to the preference for the human to be in control in analytic settings. With more capable \acro{AI} agents powered by foundation models, we believe some analytic tasks can enjoy higher levels of automation in future mixed-initiative systems.}
\added{We attribute the adoption of lower levels of automation, in part, to the preference for the human to be in control in analytic settings and the lack of access to capable artificial agents. With more capable \acro{AI} agents powered by foundation models, more analytic tasks can enjoy higher levels of automation in future mixed-initiative systems.}
The most common methods for evaluating mixed-initiative \acro{VA} systems are \leavevmode{\textbf{\color{evaluationColor}{interview studies}}} \added{(17 of 36)}
and
\leavevmode{\textbf{\color{evaluationColor}{experiments involving interaction logs}}} \added{(14 of 36)}.
Across all evaluation methods, only 8 out of 36 papers report comparative results with respect to baseline systems, whereas the remaining papers report summative results of the user studies.
Most commonly adopted principles of mixed-initiative user interfaces in visual analytics are to 
\leavevmode{\textbf{\color{miPrinciplesColor}{add significant value through automation}}} \added{(16 of 36)}
and
\leavevmode{\textbf{\color{miPrinciplesColor}{allow users to efficiently refine the results}}} \added{(19 of 36)}.
The narrow coverage of Horvitz's principles \cite{horvitz1999principles} is consistent with the narrow set of ways \acro{AI} agents currently contribute to analytic tasks and the low level of control delegated to \acro{AI} agents in present systems.

\subsection{Underexplored Areas}

\noindent\textbf{Narrow Focus on Low Levels of Automation -- }
The mixed-initiative visual analytic systems we reviewed have explored only a limited range of the design space that prioritizes human agency. As evident in our analysis, we did not observe any systems that implement fully autonomous processes and enable efficient handoffs between agents. This may be because our research discipline is intrinsically motivated to prioritize human control and leverage artificial agents for mere recommendations. However, it has also produced a potential blind spot as \acro{AI} systems grow in sophistication and there is a greater appetite for more automation. The unexplored levels of automation in the design space, \added{as highlighted by our work,} provide fruitful opportunities for our research community \added{to further explore mixed-initiative analytic settings where artificial agents are delegated more agency than seen in prior work}. 

\noindent\textbf{Evaluation  -- }
In our review, we observed \deleted{evaluation steps demonstrating} \added{evidence} that a mixed-initiative approach is better (by some metrics, especially speed) than the human performing the task alone. As \acro{AI} agents improve and become mainstream, we envision the burden of proof to shift towards showing that a mixed-initiative approach is better (given some metric\added{, including ethical considerations}) than an \acro{AI} agent completing the task alone. 
\deleted{We envision the need for} \added{By} adopting baselines in the evaluation of \deleted{our} \added{future} mixed-initiative systems, \deleted{where} we \added{we call for comparing mixed-initiative approaches} \deleted{compare} against the human and/or an \acro{AI} agent completing the task \added{alone} \deleted{on their own}. 
The broad scope of visual analytic tasks and shifting \acro{AI} capabilities make it difficult to develop robust benchmarks. However, we believe such baselines for a representative subset of well-defined tasks can further demonstrate the value of mixed-initiative interaction objectively and make visual analytics research less vulnerable to disruptions in the era of \acro{AI}.

\added{
\subsection{Opportunities for Future Work}
}

\noindent\textbf{Reusable Artificial Agents --} Prior to the emergence of foundation models~\cite{rishi2021on}, significant effort and resources were dedicated to implementing task-specific models to enable artificial agents. However, this has been obviated by foundation models and their ability to enable a myriad of uses.
As opposed to assuming brittle and rigid methods designed for narrowly scoped tasks, we envision future work to develop and evaluate modular and reusable artificial agents that can power mixed-initiative visual data analysis across various analytic environments and domain areas.

\noindent\textbf{Shared Agency --} Many reviewed mixed-initiative systems assume a fixed allocation of agency between human and artificial agents, often limiting artificial agents to making recommendations rather than taking autonomous actions. As access to advanced \acro{AI} increases, agency may shift toward more balanced human–\acro{AI} collaboration. Future mixed-initiative systems should therefore explore dynamically delegating agency between humans and artificial agents, potentially informed by factors such as risk tolerance.

\subsection{Limitations}

\deleted{Despite our rigorous process to identify and characterize mixed-initiative visual analytics systems, there are limitations to our results. First and foremost, we wish to articulate that the notion of humans and computers (whether taking the form of an \acro{AI} agent or not) is a broad and longstanding topic. With our focus being on \emph{mixed-initiative} systems only, we limited our analysis to only papers that self-identify as such. We acknowledge that there exist papers that employ mixed-initiative approaches without the use of this specific term that were left out of our Phase 2 review. With the broader ideation in Phase 1, we believe we have uncovered the core characteristics of mixed-initiative systems such as the level of automation, motivation for taking a mixed-initiative approach, and task allocation dynamics between humans and \acro{AI}. The narrow focus on mixed-initiative systems, however, hindered our ability to uncover temporal patterns on how human-\acro{AI} teaming for visual data analysis has evolved over the years. Finally, while we believe we have uncovered the core characteristics of mixed-initiative visual analytics, we do not claim for our codebook to enumerate the entire design space at a granular level. For example, while there is no item as ``faster than \acro{AI} alone'' under \leavevmode{\textbf{\color{miImpactColor}{impact of adopting a mixed-initiative approach}}}, we believe there is space for such systems in the future. }

\added{
Despite our rigorous review process, our study has several limitations. Because we focused specifically on \emph{mixed-initiative} systems, we included only papers that explicitly self-identify using this term, potentially excluding relevant work that employs mixed-initiative approaches without this label. Although the broader ideation in Phase 1 helped uncover core characteristics such as levels of automation, motivations for adopting mixed-initiative approaches, and task allocation dynamics between humans and \acro{AI}, this narrow scope limited our ability to analyze temporal trends in human-\acro{AI} teaming for visual data analysis. Finally, while our codebook captures key dimensions of mixed-initiative visual analytics, it does not enumerate the entire design space at a fine-grained level. For example, there is currently no item such as ``faster than \acro{AI} alone'' under \leavevmode{\textbf{\color{miImpactColor}{impact of adopting a mixed-initiative approach}}}, though such systems may emerge in future work.
}

\section{Conclusion}

We conducted a scoping review
\deleted{of the visual analytics literature in the past two decades}
to fundamentally characterize \emph{mixed-initiative} visual analytics systems.
In doing so, we leveraged existing taxonomies in both the visual analytics literature as well as the broader mixed-initiative systems literature. We presented our findings \added{in the form of an integrated taxonomy}, discussed \added{how existing systems populate the design space, and highlighted areas of concentration and underexplored regions.} 
\deleted{on the design space of mixed-initiative systems for visual data analysis in the era of mainstream \acro{AI} and automation.}
\deleted{We view our integrated taxonomy as a living document that can be updated as our research community makes progress in mixed-initiative visual analytics systems of the future.}
\added{Our taxonomy is not intended as a prescriptive checklist, but as an analytical lens that captures the current state of the field and supports more deliberate design decisions in future systems.}

\section*{Acknowledgments}
We thank Chad A.\ Steed for many inspiring discussions.
Research sponsored, in part, by the Laboratory Directed Research and Development Program of Oak Ridge National Laboratory, managed by UT-Battelle, LLC, for the US Department of Energy.

\printbibliography

\end{document}